\documentclass[conference,9pt]{IEEEtran}
\IEEEoverridecommandlockouts
\usepackage{spconf}
\usepackage{url}
\usepackage{subcaption} 
\usepackage{amsmath}
\usepackage{amssymb}
\usepackage{xcolor}
\usepackage{multirow}
\usepackage{etoolbox,balance}

\usepackage{cite}

\usepackage{balance}
\usepackage{pifont}
\usepackage{stmaryrd}
\usepackage{optidef}
\usepackage{mathrsfs}
\usepackage{amsfonts}  
\usepackage[bookmarks=false]{hyperref}

\usepackage{graphicx}
\usepackage{algorithm}
\usepackage{cuted}

\usepackage{algorithm}
\usepackage{algpseudocode}

\usepackage{booktabs}
\usepackage{dsfont}
\usepackage{esint}

\patchcmd{\thebibliography}{\section*{\refname}}{}{}{}

\title{\vspace{-2.5em} Signal Processing Challenges in Automotive Radar}
\name{Sandeep Rao$^{\dagger}$, Rajan Narasimha$^{\ddagger}$, and Shunqiao Sun$^{\S}$
\thanks{\hspace{-1.0em}The work of S.\ Sun was supported in part by U.S. National Science Foundation (NSF) under Grants CCF-2153386 and ECCS-2340029.}}
\address{$^{\dagger}$Texas Instruments, Bangalore, India, Email: s-rao@ti.com\\
$^{\ddagger}$Texas Instruments, Dallas, TX, USA, Email: ln.rajan@ti.com \\
$^\S$The University of Alabama, Tuscaloosa, AL, USA, Email: shunqiao.sun@ua.edu }

\begin{document}
%
\maketitle

\begin{abstract}
As automotive radars continue to proliferate, there is a continuous need for improved performance and several critical problems that need to be solved. All of this is driving research across industry and academia. This paper is an overview of research areas that are centered around signal processing. We discuss opportunities in the area of modulation schemes, interference avoidance, spatial resolution enhancement and application of deep learning. A rich list of references is  provided. This paper should serve as a useful starting point for signal processing practitioners looking to work in the area of automotive radars.
\end{abstract}

\begin{keywords}
FMCW Radar, Automotive Radar, Interference Mitigation, Spatial Resolution 
\end{keywords} 
\section{Introduction}
Millimeter wave radars are a critical part of the solution for assisted and autonomous driving due to their robustness in diverse weather, performance that is independent of lighting  and ability to accurately measure and track the range and velocity of targets \cite{SUN_SPM_Feature_Article_2020,sun20214d}. Further, advances in CMOS technology have significantly improved the integration level so that transceivers, ADCs and signal processing units can be integrated into a single die. This has driven down the cost and size of radar sensors. However, to meet the requirement of Level 4 and 5 autonomous driving there are several challenges that must be overcome, the most pressing of which are listed below. 

The spatial resolution of the current generation of radars is limited and needs to be improved. As the number of radars on the road increase, reliably detecting and mitigating mutual interference becomes important. The next generation of radars will move beyond target detection and tracking to also perform target classification. Data fusion across multiple radars on the car, and across multiple sensing modalities (e.g., camera + radar) is also becoming increasingly relevant.

While almost all radars on the road today use a modulation scheme known as Frequency Modulated Continuous Wave (FMCW), there is work on alternate modulations that may be more suited for future radar requirements: increased spatial resolution, longer range, higher measurable maximum velocity, robustness to interference – all this while maintaining low cost and low power. 

This paper provides a quick introduction to radars, the current state of art and unsolved challenges that are of interest to the signal processing community.  We hope that this paper serves as a useful starting point for signal processing practitioners looking to work in the area of automotive radars. The paper is organized as follows. The Section II provides a brief introduction to the PHY layer of radar- the underlying modulation schemes  and associated signal processing. We also highlight the challenges in finding an optimal modulation scheme. Section III addresses the problem of interference mitigation. Section IV addresses ongoing work to enhance spatial resolution. Section V  briefly outlines deep learning approaches in automotive radar.

\section{MODULATION TECHNOLOGIES}\label{sigm}
A radar transmits a signal, and based on the reflected signals estimates the range and relative velocity of targets in the scene. Additionally, with multiple receive and transmit antennas, the radar also estimates the angle of arrival of  targets. 

\subsection{Modulation Techniques}
\begin{figure}[h]
    \centering
        \includegraphics[width=3.30 in]{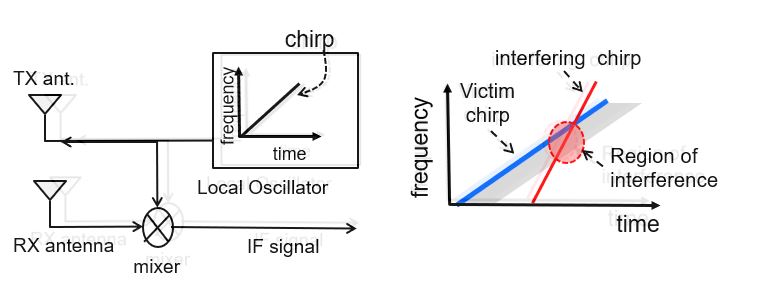} 
    \caption{FMCW block diagram (left), Mutual interference (right)}
    \label{fig:one}
\end{figure}

``Frequency Modulated Continuous Wave (FMCW)'' is the modulation technology currently used in almost all automotive radars \cite{TI_fundamental_mmW_2020, TI_mmW_Sensing_video_2024}. The fundamental signal in FMCW radar is a chirp. A chirp is a signal whose frequency changes linearly with time. The Local Oscillator (LO) generates a chirp that is transmitted by an antenna (in automotive radar this chirp might ramp from 77GHz to 79GHz). The received signal (reflected from the scene in front of the radar) is mixed with the transmit signal to create an intermediate frequency (IF) signal (Fig. \ref{fig:one}(left)).  It can be shown that the frequency of the IF signal is proportional to the round trip delay of the received signal. Thus, in the presence of multiple reflections, the received signal consists of multiple frequency components. The received IF signal is digitized and the digitized data sent to a processor that performs an FFT, to produce a ‘range-profile’ where the location of peaks directly corresponding 
to the distance of targets relative to the radar.

Phase Modulated Continuous Wave (PMCW) is another modulation technique, where a pseudorandom noise (PN) coded binary signal replaces the chirp as a fundamental unit of transmission \cite{PMCW_RadarConf_2016}. A correlation is performed at the receiver, with peaks corresponding to the round trip delay of various reflections. So a correlation replaces the FFT as the operation which produces the range profile. 

Recently there has also been some discussion of Orthogonal Frequency Division Multiplexing (OFDM) as an alternate modulation scheme \cite{MIMO_OFDM_MTT_2018}. The  motivation is a common modulation technology for radar sensing and vehicle to vehicle communication. In OFDM radar the simultaneous transmission of multiple tones forms the fundamental transmit signal. Optionally, each of the tones can be multiplied by a data symbol. At the receiver the process of obtaining the range-profile involves an FFT on the received ADC samples, wiping out the modulated data, followed by an inverse FFT.

\subsection{Radar Signal Processing}
While the process of the obtaining the range-profile depends on the choice of the modulation scheme, subsequent processing (velocity and angle estimation) is largely independent of the underlying physical layer. For velocity estimation, multiple equi-spaced replicas of the `fundamental signal' are transmitted in a unit called a frame. Each transmission of the fundamental signal produces a range-profile. Relative motion between the radar and the target means that subsequent transmissions have slightly different round trip delays to the same target. These differences are very small (in the order of a fraction of wavelength of the transmit signal). However, these show up as phase changes in the range-profile (across subsequent transmissions of the fundamental signal).  FFT’s performed across range-profiles convert these phase changes to peaks in an FFT spectrum (the location of the peaks being directly proportional to the relative velocity of the target).

Estimating the angle of arrival of targets requires multiple transmit / receive antennas.  Different transmit/receive antenna pairs have slightly different round trip delays to the same target, the path delays dependent on the angle of arrival of the received signal. Again, phase differences between these different paths are used to estimate the angle of arrival of the target. 

Many performance parameters of radar are independent of the specific modulation scheme. For e.g., the range resolution is inversely proportional to the RF bandwidth ($f_{\rm BRF}$) (and given by $c/2f_{\rm BRF}$) and the velocity resolution is inversely proportional to the frame length $T_f$ (given by $\lambda/2T_f$), where $c$ is the speed of light and $\lambda$ is the wavelength. The maximum unambiguously measurable velocity is inversely proportional to the spacing ($T_c$) between adjacent ``fundamental signals'' in a frame ($\pm \lambda/4T_c$).  An overview of radar signal processing is provided in \cite{TI_fundamental_mmW_2020, TI_mmW_Sensing_video_2024}.

\subsection{Comparison of Modulation Technologies}
Despite these commonalities, there are differences between  modulation schemes in hardware complexity and associated system performance. FMCW uses a constant envelope transmit signal, which eases requirements on the power amplifier (PA). The IF bandwidth requirement of an FMCW radar is orders of magnitude less than the RF bandwidth. So, while the RF bandwidth is typically of the order of a few GHz (corresponding to range resolution of a few cms), an FMCW radar requires an IF bandwidth of only a few tens of MHz. In contrast, PMCW and OFDM based radars require ADC’s that sample at a rate that’s commensurate with the bandwidth of the RF signal (for high resolution radars this is of the order of Gsps).  A common problem in radars is the saturation of the receiver  due to the on-board coupling with the transmitter. In FMCW radar, the coupling signal shows up as a very low frequency component in the IF signal. Thus, a simple high pass filter prior to the ADC can prevent this signal from flooding the ADC. In contrast removing the coupling signal prior to digitization is a much more challenging problem in OFDM/PMCW radar. The easing of requirements on the hardware (e.g. ADC, PA) is the primary reason for the popularity of FMCW radar. The primary advantage of PMCW (and OFDM) radar is that (for the same RF bandwidth) they tend to allow significantly larger repetition rate in the “fundamental signal”, which translates to larger native maximum velocity. 
\section{INTERFERENCE MITIGATION}
Mutual interference for FMCW radars has been well studied \cite{Alland_Interference_SPM_2019,Rao_Interference_RadarConf_2020}. The small IF bandwidth of an FMCW radar (compared to its RF bandwidth) translates to the interferer being localized to small portions of the signal. This is illustrated in Fig. \ref{fig:one} (right)) which depicts a victim chirp (blue) and interfering chirp (red). As explained in \cite{Rao_Interference_RadarConf_2020}, due to IF filtering, only a signal that enters the grey zone causes interference. Thus, an interfering chirp with a significantly different slope will corrupt only a small segment of the received signal.  An interfering chirp with a similar slope as the victim chirp can corrupt almost the entire victim chirp. However, techniques such as randomization of the inter-chirp duration, can ensure that only few chirps in a frame are thus corrupted. Interference mitigation strategies can be broadly divided into two categories: reactive and proactive.

\subsection{Reactive Mitigation}
Reactive interference mitigation schemes rely on the fact that interference events in FMCW radars are localized, corrupting a small segment of the received signal of specific chirps, or alternatively a small set of entire chirps in a frame.  They typically follow a two-step process (a) localization of interference and (b) reconstruction of the excised portion of the signal.

{\bf Interference Localization:} The simplest interference localization scheme in FMCW radar is to monitor the signal strength of the received IF signal. Segments of the signal with a statistically higher signal value compared to the rest of the received signal are identified as interference affected. This exploits the fact that in many cases,  the interferer signal is higher in energy than the received signal from a target. This is because the received signal strength from a target at distance $R$ is proportional to $R^{-4}$ (due to the round-trip delay from the radar to the target and back, each path contributing $R^{-2}$). The interfering signal on the other hand only has to traverse a single path (from interfering radar to victim radar) thus experiencing a path loss of only $R^{-2}$. More sophisticated schemes use a high pass filtered version of the received IF signal to detect interference \cite{Rao_interference_RadarConf_2021}. This works because higher frequencies in the IF band (corresponding to farther targets) will have lower signal strength making it easier to detect weaker interferers. 

{\bf Signal Reconstruction:} Once the interference affected part of the received signal has been identified and excised, the next step is reconstructing the missing portion of the signal. Simple schemes such as filling the missing portion with zeros or linear interpolation across missing samples are commonly used. More sophisticated signal reconstruction techniques that have been explored include use of autoregressive models \cite{Autoregressive_interference_Sensors_2021}, recursive reconstruction techniques such IMAT \cite{Interference_EuRAD_2017} and  OMP \cite{interference_IMCIM_2019}. A good comparison of various reconstruction techniques is in \cite{Yash_interference_ICASSP2024}. A different approach that  estimates the parameters of the interferer (assuming FMCW) and then subtracts the interfering component from the corrupted signal is proposed in \cite{interference_IMCIM_2015}. 

``Sense and Avoid'' is another reactive scheme \cite{TI_interference_note_2022}. It involves  sensing the spectrum for interference and then switching to a time slot/frequency band that is free of interference. \cite{int_sriram} explores the use of frequency bands outside the IF band of interest to  monitor interferers. Interference localization and reconstruction schemes are particularly suited for FMCW radars – and do not work as well for OFDM or PMCW radars. \cite{int_bourdoux},\cite{int_carvajal}  compares various modulation schemes in the presence of interference.

\subsection{Pro-active Methods}
{\bf Dithering of TX Signal Parameters:} Here certain parameters of the transmit signal are randomized \cite{TI_interference_note_2022} E.g. the phase of chirps can be varied across a frame using a sequence unique to each radar. This ensures that the interfering signal is spread out during Doppler processing thus preventing the appearance of ghost targets.  

{\bf Co-ordination:} This involves coordinating radar transmissions so that they occupy non-overlapping time and frequency resources. This co-ordination is achieved via communication (either via a dedicated channel e.g. V2X, or time multiplexing  radar and communication over a common channel).

Despite a large existing body of work, interference mitigation is a work in progress. Reconstruction and dithering schemes can suffer from increased noise floor/ghost targets. Some interference mitigation algorithms can be very compute heavy and not suitable for low-cost radars. Co-ordination among radars requires a significant effort towards standardization. With increase in both the number of cars with radars and the number of radars per car, the problem of interference  will become more pressing in the future.  

\section{ENHANCING SPATIAL RESOLUTION}\label{perf}
Radars have great range and velocity resolution. The range resolution depends on the RF bandwidth spanned by the TX signal. A bandwidth of 4GHz (maximum permissible by regulation) translates to a range resolution of 4cm.  The velocity resolution depends only on the frame time (the larger the better). Today’s radars can easily achieve velocity resolution of a fraction of a meter/sec. However, the spatial resolution of the current generation of radars is substantially lesser than what is achievable by lidar or camera. 

The most obvious way to increase the spatial resolution is to increase the number of antennas. This comes at the expense of area and cost. Here there are two options. The first  is to cascade multiple radar chips (each with a limited number of transceivers) so that all these devices are synchronized and can operate with phase coherency to effectively create a large phase array \cite{TI_cascade}. The second option is to integrate a large number of transmit and receive channels on a single die to achieve high angle resolution \cite{imaging_uhnder}. The former solution has the advantage of scalability: each application can design a phase array to meet a specific angle resolution requirement by cascading different number of devices as needed. However, it could face challenges of synchronization across the devices. On the other hand, the latter option features high level of integration with inherent synchronization, but loses flexibility and faces challenges of heat dissipation that significantly impacts the performance stability across temperatures. Another approach to improve spatial resolution is to employ specialized algorithms to extract the maximum spatial resolution out of a given antenna array. This is explored in the sub-section below.

\subsection{High Angle Resolution Algorithms and Sparse Arrays}
The fundamental challenge for automotive radar super-resolution is to achieve high angular resolution ($0.5^\circ-1^\circ$) in a dynamic environment, limiting the availability of snapshots to just one in most situations \cite{SUN_SPM_Feature_Article_2020}. The challenge is further compounded by the presence of potentially large number of sources with dynamic ranges as large as 30dB and signal-to-noise ratio in the 15-20 dB range. The uniform linear array (ULA) size required to achieve such high resolution with Bartlett beamforming is 200 elements for the $0.5^\circ$ resolution target. The goal of achieving high resolution with a practical number of antennas has led researchers to investigate several approaches for super-resolution including sparse arrays. The landscape of arrays and algorithms for super-resolution is vast, owing to the large diversity in problem settings (SNR, snapshots, source count vs sensor count and model priors), optimization cost functions (direct measurements or function of measurements such as subspace, covariance \cite{VibergSurvey_2023}), optimization approaches (greedy, convex, non-convex) and attempts to trade-off performance and complexity.  

Traditional super-resolution techniques such as MUSIC assume the availability of a large number of uncorrelated snapshots, not realistic for automotive radar. Single snapshot MUSIC has been studied for ULA \cite{liao2014musicsinglesnapshotspectralestimation}, but not applied to sparse arrays. The class of deterministic arrays such as minimum redundancy arrays (MRAs), nested arrays \cite{ppalNested_2010} and co-prime arrays, combined with angle estimation performed on the difference coarray via spatial smoothing or direct augmentation \cite{pillai_augment1985} has been proposed as a low-complexity super-resolution method. While this approach has the key advantage of identifying more sources than sensors, it is limited by the non-availability of larger number of uncorrelated snapshots. The above limitations have led researchers to investigate the body of sparse-recovery methods, inspired by developments in compressive sensing (CS). As a starting point, parametric methods such as deterministic or stochastic maximum likelihood (DML, SML, respectively \cite{Van_Trees}), applicable to arbitrary arrays, require solving a non-linear least squares problem given the true number of sources. The ML cost function is non-convex and determining the global optimum requires exhaustive search over the grid that grows exponentially with the number of sources. Therefore, this approach is usually not viable for estimating more than two sources. Sparse recovery for direction of arrival(DoA) estimation has evolved along some key trajectories to address this complexity bottleneck of ML by exploring trade-offs between complexity and performance. 

On the one hand are the gridded methods inspired by CS literature that partition the FoV into a discrete grid and assume that the true source is present at a subset of grid locations. At the lower end of the complexity scale of the gridded methods are greedy (suboptimal) approaches that take the sequential approach of estimating one source in each iteration. Prime examples of these include matching pursuit (MP), orthogonal matching pursuit (OMP) and orthogonal least squares (OLS) \cite{ziskind_ml1988}. These methods sacrifice performance for complexity, suffer from estimation biases arising from estimating one source at a time. They differ from each other with regard to how the source amplitudes are updated in each iteration. At the next level of hierarchy are the semi-parametric methods that optimize a cost function that balances data fidelity and sparsity ($\ell_0$ norm) via a regularization parameter. The $\ell_0$ norm (non-convex) is replaced by the $\ell_1$ norm which is convex. Basis pursuit (BP) and Basis Pursuit Denoising (BPDN) \cite{malioutov_bpdn2005} are key variants of this approach with a convex cost function and established performance guarantees when the array satisfies restricted isometry property (RIP) which is hard to establish. This method is aided by advances in convex optimization resulting in polynomial time complexity. While the use of convex optimization solver (semi-definite programming SDPT/self-dual minimization SeDuMi) can incur high complexity when the problem dimension is large, fast $\ell_1$ methods (\cite{yang_fastL12010} iterative shrinkage thresholding, ISTA, Nestorov's method NESTA etc.) have been presented in literature to speed up the computation. FOCUSS \cite{bdrao_focuss1997} is a higher performance but non-convex variant when the $\ell_1$ norm based sparsity inducing term above is replaced by a $\ell_q$ norm ($0 < q < 1$). The difficulty in tuning the hyperparameter for BPDN/FOCUSS has lead to the development of methods that assume a parameterized distribution for the amplitude priors and apply the MAP criterion to iteratively estimate the amplitudes and hyperparameters from the data. Some key algorithms taking this approach include sparse learning via iterative minimization (SLIM), sparse Bayesian learning (SBL, \cite{bdrao_SBL2004}) and Bayesian Linear Regression with Cauchy prior (BLRC) \cite{BLRC_2023}. Another recent parameter-free approach is the  sparse iterative covariance estimation (SPICE \cite{stoica_spice2011}) algorithm that optimizes a convex cost function based on covariance matching. 

On the other hand, the gridless methods are based on recent advances in matrix completion. The first stage of these methods is to complete the Hankel matrix corresponding to the full ULA via partially observed entries from the sparse array using matrix completion techniques \cite{ychie_matcomp2014,sun20214d}. Once the Hankel matrix for the full ULA is estimated, the directions of arrival can be recovered by classical subspace methods such as MUSIC. 

In summary, the state-of-the-art consists of various array and algorithm approaches offering performance vs complexity trade-offs. No one method has been established to resolve all the challenges arising in automotive radar. This remains an evolving field with the goal of improving accuracy and identifying multiple sources with large dynamic range at reduced complexity. Some directions of research include distributed arrays, exploring arrays with non-integer spacings, automatic model order selection and data driven approaches to overcome the limitation of model-based approaches. The development of h/w acceleration methods for the class of sparse recovery methods is another key direction.

\section{APPLICATION OF DEEP LEARNING}
Radars have very good velocity resolution. Hence the use of deep learning (DL) to classify targets based on their motion signature, i.e., micro-Doppler, is a well studied topic with \cite{dl_sevgi} providing an excellent overview. Results in \cite{dl_vru} suggest that similar approaches would work in the context of automotive radar to classify vulnerable road users such as pedestrians and cyclists. 

With its robustness to weather conditions, independence from ambient light, superior range and velocity resolution and low cost,  radar is a perfect complement to camera. 
DL has been used for fusion  with other sensing modalities \cite{dl_fusion_nobis}.  \cite{dl_review} is an overview of DL in senor fusion, target detection and occupancy grid mapping. 

Interestingly DL is also been used in domains that have traditionally relied solely on signal processing. For e.g, data-driven deep learning  approaches for DoA estimation have gained significant traction \cite{papageorgiou2021deep,fuchs2022machine,feintuch2023neural}.   Authors have reported improved resolution and robustness in low signal-to-noise ratio scenarios \cite{papageorgiou2021deep}.  DL has also been applied for interference mitigation \cite{dl_int_overdevest,dl_int_jiang}. 

However, deep learning techniques are predominantly data-driven and often lack interpretability. Model-based deep learning methods \cite{shlezinger2023model} aim to bridge this gap by combining the strengths of  mathematical and physics models with data-driven machine learning, which has shown high-resolution performance and better generalization capabilities \cite{Zheng_Sensors_DOA_model_based_DL_2024}. One such example is algorithm unrolling \cite{Monga_SPM_Unrolling_2021} which replaces the iterations of an optimization algorithm with  neural layers, such as learned ISTA (LISTA) \cite{gregor2010learning}.   

\section{Conclusions}
Radars can accurately sense in dimensions of range and velocity, are inexpensive and are robust to weather conditions (sun light, fog, rain) and are thus receiving lot of attention in both industry and academia.  This overview has covered areas of research in automotive radars that would be of interest to signal processing practitioners. We discussed two key areas:  interference mitigation and spatial resolution enhancement both of which can benefit from signal processing algorithms, approaches blended with machine learning  and innovations at the PHY layer. Higher layer capabilities such as detection, tracking and fusion of radar with other sensing modalities are also important research areas.

\bibliographystyle{IEEEtran}

\bibliography{refs}
\end{document}